\documentclass[pra,twocolumn,groupedaddress,floatfix]{revtex4}

\usepackage[dvips]{graphicx}%
\usepackage{amsmath}

\usepackage{bm}

\begin{document}
\title{Efficient-phase-encoding protocols for continuous-variable quantum key distribution using coherent states and postselection}
\author{Ryo Namiki}
\email[Electric address: ]{namiki@qo.phys.gakushuin.ac.jp}
\affiliation{CREST Research Team for Photonic Quantum Information, Division of Materials Physics, Department of Materials Engineering Science, Graduate school of Engineering Science, Osaka University, Toyonaka, Osaka 560-8531, Japan}

\author{Takuya Hirano }
\affiliation{Department of Physics, Gakushuin University, Mejiro 1-5-1, Toshima-ku, Tokyo 171-8588, Japan}

\date{\today}
\begin{abstract}
We propose efficient-phase-encoding protocols for continuous-variable quantum key distribution using coherent states and postselection. 
By these phase encodings, the probability of basis mismatch is reduced and total efficiency is increased. We also propose mixed-state protocols by omitting a part of classical communication steps in the efficient-phase-encoding protocols. The omission implies a reduction of information to an eavesdropper and possibly enhances the security of the protocols. We investigate the security of the protocols against individual beam splitting attack.
\end{abstract}

\maketitle

\small
\section{Introduction}

In quantum key distribution (QKD), two distant parties, Alice and
Bob share a secret key exploiting quantum channel and classical communication. By the laws of quantum physics the key can be proved to be secure against an eavesdropper (Eve) who has advanced technologies \cite{rmp74}.  
 Many novel and modified QKD protocols have been proposed based on various theoretical and experimental aspects. For example, in the standard weak-coherent-state (WCS) implementation of the BB84 protocol, the transmission distance of QKD is limited by the photon-number-splitting (PNS) attack \cite{pns}. To defeat this limitation, modifications of the implementation have been proposed. An elegant one is the SARG protocol which changes the classical communication step in the implementation \cite{SARG}. Another one is the decoy-state protocols that utilize fake signals to restrict possiblility of the PNS attack \cite{DS1,DS2}.

A different type of WCS implementation has been proposed combining the idea of the balanced homodyne detection of an weak signal field with a strong field in the phase-encoding interferometric implementation of the BB84 protocol \cite{hirano}. This proposal also includes the idea of postselection in continuous variable (CV) QKD. 
Following BB84, the protocol has inefficiency of basis mismatch associated with the random basis exchange, the choice of the quadratures. However, there are several CV QKD protocols that have no such inefficiency \cite{coherent,postsel,coherentR,no-switch}. In relation with the implementation, those protocols employ the amplitude modulations in addition to the phase modulation, and the signals are not necessarily WCS. 
Although CV QKD protocols are free from the limitation by the PNS attack, another practical limitation is given by the classical teleportation attack \cite{namiki2,namiki3}.

In this paper we propose efficient phase-encoding protocols for CV QKD using balanced homodyne detection and postselection those have better efficiency than the original one \cite{hirano} without significant changes in experimental setup.
We also propose mixed-state protocols by omitting the part of classical communication steps. 

In Sec. II, we review the original protocol and provide a basic notation. In Sec. III, we present modified protocols. In Sec. IV, we investigate the security against individual beam splitting attack. Sec. V is the conclusion and remarks.

\section{original four-coherent-state postselection protocol}%

\begin{figure}[htbp]
\includegraphics[width=6cm]{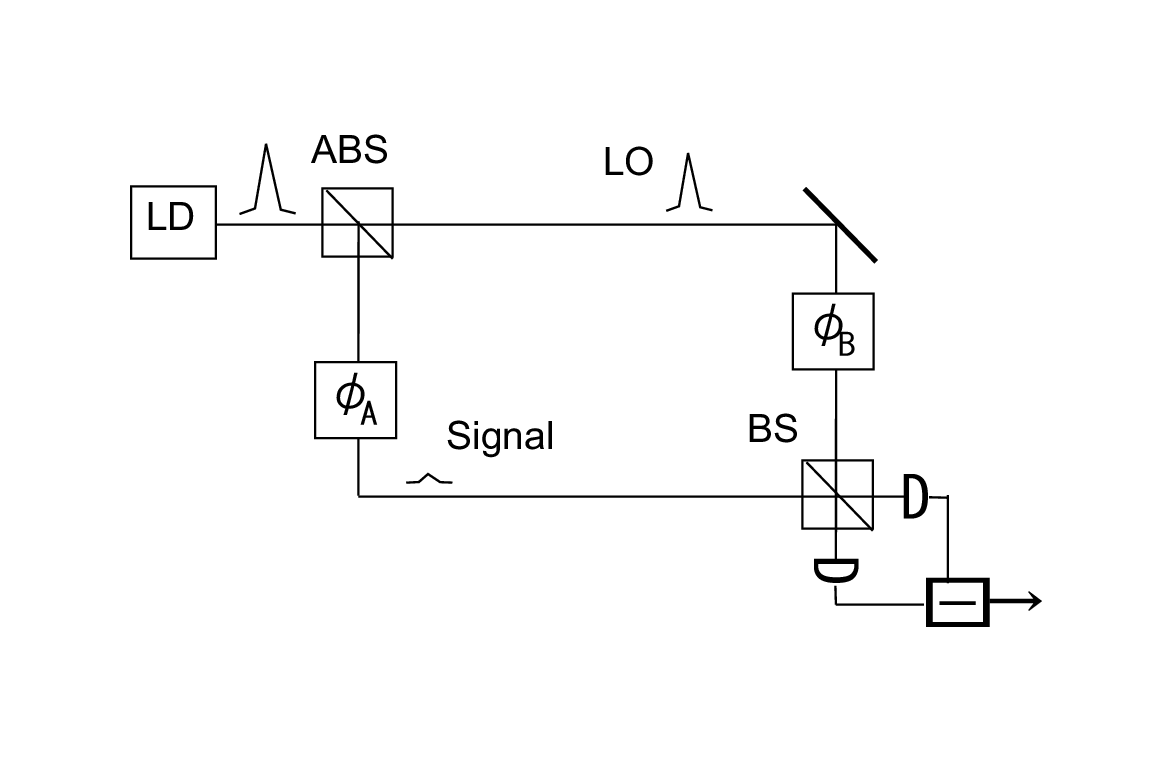}
\caption{The pulse emitted from a light source (LD) is split into a strong local oscillator (LO) and a weak signal field by an asymmetric beam splitter (ABS). Alice applies her phase shift $\phi_A$ to the signal and Bob applies his phase shift $\phi_B$ to the LO. The signal and LO interfere at the 50:50 beam splitter (BS). The difference between the photon numbers of the output pulses of the BS is observed. \label{if1}}
\end{figure}

 The original four-state postselection protocol (O4) \cite{hirano,namiki1} is based on the phase-encoding interferometric implementation of QKD and the balanced homodyne detection of a weak signal field with a strong field as in FIG. \ref{if1}.

 Alice sends the coherent state $| \alpha e ^{i \phi_A} \rangle $ with $\alpha > 0$ by randomly choosing her phase modulation $\phi_A$ from a set $\{ 0, {\pi/2},{  \pi },{3 \pi/2}  \}$.
Bob measures the quadratures $\hat x(\phi_B) \equiv \hat x_1 \cos \phi_B + \hat x_2 \sin \phi_B $ by  randomly choosing his phase modulation $\phi _B $ from a set $\{ 0, {\pi/2}   \}$ where $\hat x_1 \equiv \frac{\hat a + \hat a ^\dagger }{2}  $ and $\hat x_2 \equiv \frac{\hat a - \hat a ^\dagger }{2i}  $.
 Bob's measurement is characterized by the quadrature distribution of the coherent state $| \langle x_{\phi_B}| \alpha e^{i\phi_A } \rangle |^2$ where $ \langle x_{\phi_B}|$ is the eigenbra of $\hat x(\phi_B)$ with the eigenvalue $x_{\phi_B}$. For a simpler notation we define 
\begin{eqnarray}P(x,\alpha, \phi)  &\equiv & \sqrt\frac{2}{\pi}\exp \left\{ -2 (x - \alpha \cos  \phi )^2\right\}  .
\end{eqnarray} Then, we can write
\begin{eqnarray}| \langle x_{\phi_B}| \alpha e^{i\phi_A } \rangle |^2 = P(x_{\phi_B},\alpha ,{ \phi_A -\phi_B}).
\end{eqnarray}

\begin{figure}[htbp]
\includegraphics[width=5cm]{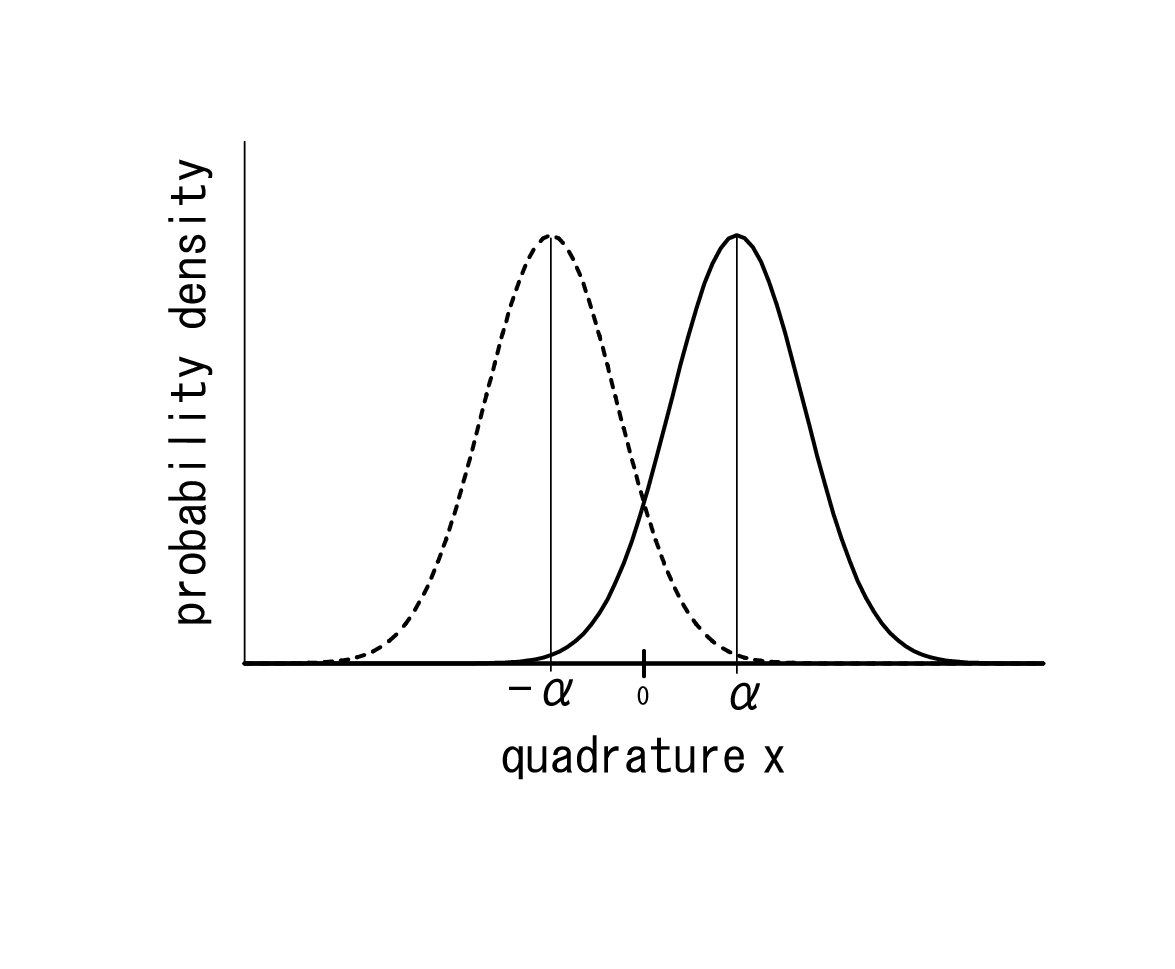}
\caption{The quadrature distributions of correct-basis choice are shown. The solid curve peaked at $x = \alpha $ corresponds to the signal of Alice's bit ``1'' and the dotted line peaked at $x = -\alpha $ corresponds to the signal of Alice's bit ``0''. \label{dist} }
\end{figure}

 After the transmission, Bob informs Alice of his phase shift $\phi_B$. If $ |\phi_A-\phi_B | = \{0, \pi\} $, Bob's distribution is one of the Gaussian distributions centered either at $\pm \alpha $ as in FIG. \ref{dist}. We call such a combination of $(\phi _A,\phi _B)$ the \textit{correct basis} and the quadrature distribution is given by
\begin{eqnarray}
P_B(x, \alpha )= \frac{1}{2}\left(P(x, \alpha , 0 )+ P(x,\alpha , \pi )\right), 
\end{eqnarray} where $x$ corresponds to Bob's  measurement outcome.

 For the correct-basis case, the signal can transfer bit information from Alice to Bob by the following manner.  Alice encodes her bit according to the combination of $(\phi _A,\phi _B)$ as in Table \ref{table1}. Bob decodes the bit value according to his outcome $x$: If $x \ge  0$ his bit value is ``1'' otherwise his bit value is ``0''.
The bit error rate (BER) conditioned on the absolute value $|x|$ is given by
\begin{eqnarray} q_B(x, \sqrt \eta \alpha  ) &=&  \frac{ P(|x|, \sqrt \eta  \alpha , \pi )}{P(|x|, \sqrt \eta \alpha , 0 )+ P(|x|, \sqrt \eta  \alpha , \pi )} .
\end{eqnarray} where we assume the lossy channel with the line transmission $  \eta $, $( 0 <\eta \le 1)$. 
Since the quadrature distributions are spread and overlapped, Bob's positive (negative) quadrature result needs not correspond to Alice's bit ``1" (``0'') and Bob's decoding has inherently finite errors. However, $q_B$ is less than $\frac{1}{2}$ if $x \neq 0$ and non-zero information is transferred.
By selecting the data according to the value of $x$, Alice and Bob can discard the erroneous portion. This process is called postselection.

If $(|\phi_A-\phi_B | \mod \pi )= \pi/2  $, the signal is called the \textit{wrong basis}. In this case, they cannot share the bit information because Bob's quadrature distribution is the same for such a combination of $\phi_A$ and $\phi_B$, and the wrong-basis signal is discarded. 
 
\begin{figure}[htb]
\begin{center}
\includegraphics[width=5cm]{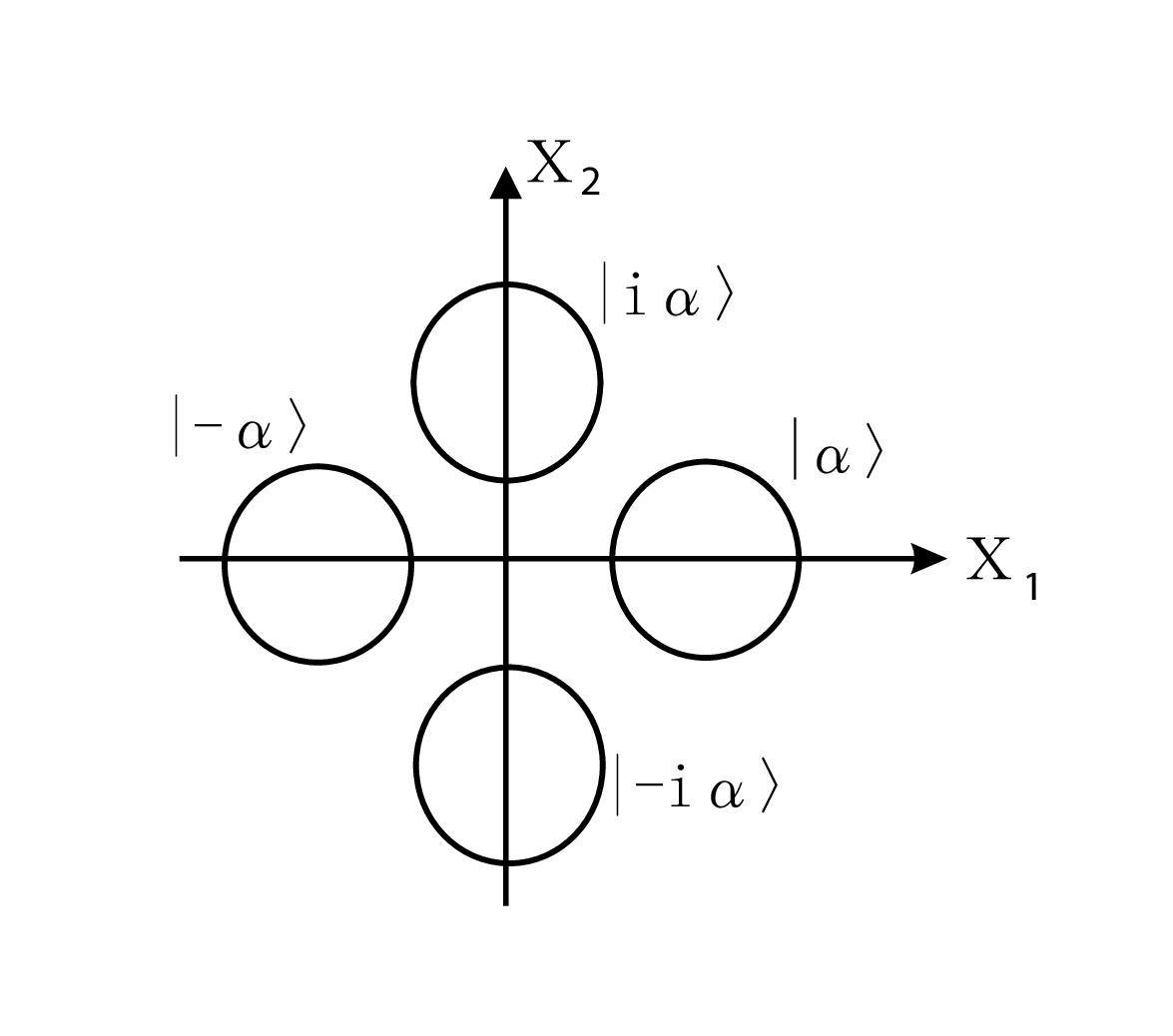}
\caption{Phase-space picture of the original four-state protocol\label{ps-o4}}\end{center}
\end{figure}
\begingroup
\squeezetable
\begin{table}[htbp]
 \caption{Alice's bit encoding of the original four-state protocol  \label{table1} }%
  \begin{tabular}{c||c|c|c|c|c|c|c|c|}
\hline
    $\phi_A $ &0 &0   &  $\pi/ 2$  &   $\pi/ 2$  &   $\pi$ &  $\pi$ &  $3\pi/ 2$ & $3\pi/ 2$    \\
    \hline
     $\phi_B $  &  0  &  $\pi /2$  &  0  &  $\pi /2$ & 0 & $\pi /2$ & 0 & $\pi /2$ \\
    \hline
     $\langle \hat x \rangle $  &  $\alpha $  &  0  & 0   &  $\alpha $  &  $-\alpha $& 0 & 0  &  $-\alpha $  \\
    \hline
  $A$  &  1  &   &    &  1 &  0&   &   & 0    \\
    \hline
  \end{tabular}
\end{table} 
\endgroup

We define the efficiency $P_e$ of the protocol as the probability that the signal becomes correct basis. In this O4 protocol one-half of signals are discarded and 
\begin{equation}
P_e = \frac{1}{2}. 
\end{equation}
In the modified protocols, this quantity will be improved.
 
Alice's bit encoding of the O4 protocol associated with $\phi_A $, $\phi_B $, and Bob's mean value of quadratures $\langle \hat x \rangle \equiv \langle \alpha e^{i\phi_A} |\hat x(\phi_B)  |\alpha e^{i\phi_A} \rangle$ is summarized in Table I where $A$ represents Alice's bit. For the combination of wrong basis, $\langle \hat x \rangle$ is zero and $A$ is set to be blank. The probability of wrong basis is determined by the number of blanks in the row $A$.
A phase-space description is shown in FIG. \ref{ps-o4}. %

Towards the modification, the essential point of performing the postselection is that Bob's distribution takes the form like FIG. \ref{dist}. As we will show, the state configuration of FIG.\ref{ps-o4} and the random choice of a conjugate-quadrature pair is not necessary.

For a convenience, we set \textit{Alice's bit encoding rule}: The combination of $(\phi_A, \phi_B)$ that leads to $\langle \hat x \rangle =  0$ is for wrong basis,   
 $\langle \hat x \rangle = \alpha$ is for bit ``1'' and $\langle \hat x \rangle = -\alpha$ is for bit ``0''. Note that Table I is consistent with this rule. We use this rule repeatedly in the following section.

The security of the protocol can be related to the uncertainty relation of the quadratures, \begin{eqnarray} (\Delta x_1)^2 (\Delta x_2)^2 \ge \frac{1}{16}.  \end{eqnarray}
Let us consider the ideal case, i.e., the channel and detector are lossless and noiseless. In such a case Bob can measure the mean and variance of $\hat x_1$ and $\hat x_2$ for each of the four coherent states and confirm that each of the states is in the minimum uncertainty state with $(\Delta x_1)^2=(\Delta x_2)^2 = \frac{1}{4}$. Since such a minimum uncertainty state is a pure coherent state, this implies that a set of non-orthogonal states transfers without any disturbance. Thus, there is no Eve's intervention if the variances and mean values of quadratures have no changes, and the minimum uncertainty ensures the security. %

\section{Efficient postselection protocols}
 In this section we present several modified protocols. 

\subsection{ Three-state protocol }
\begin{figure}[htbp]
\begin{center}
\includegraphics[width=5cm]{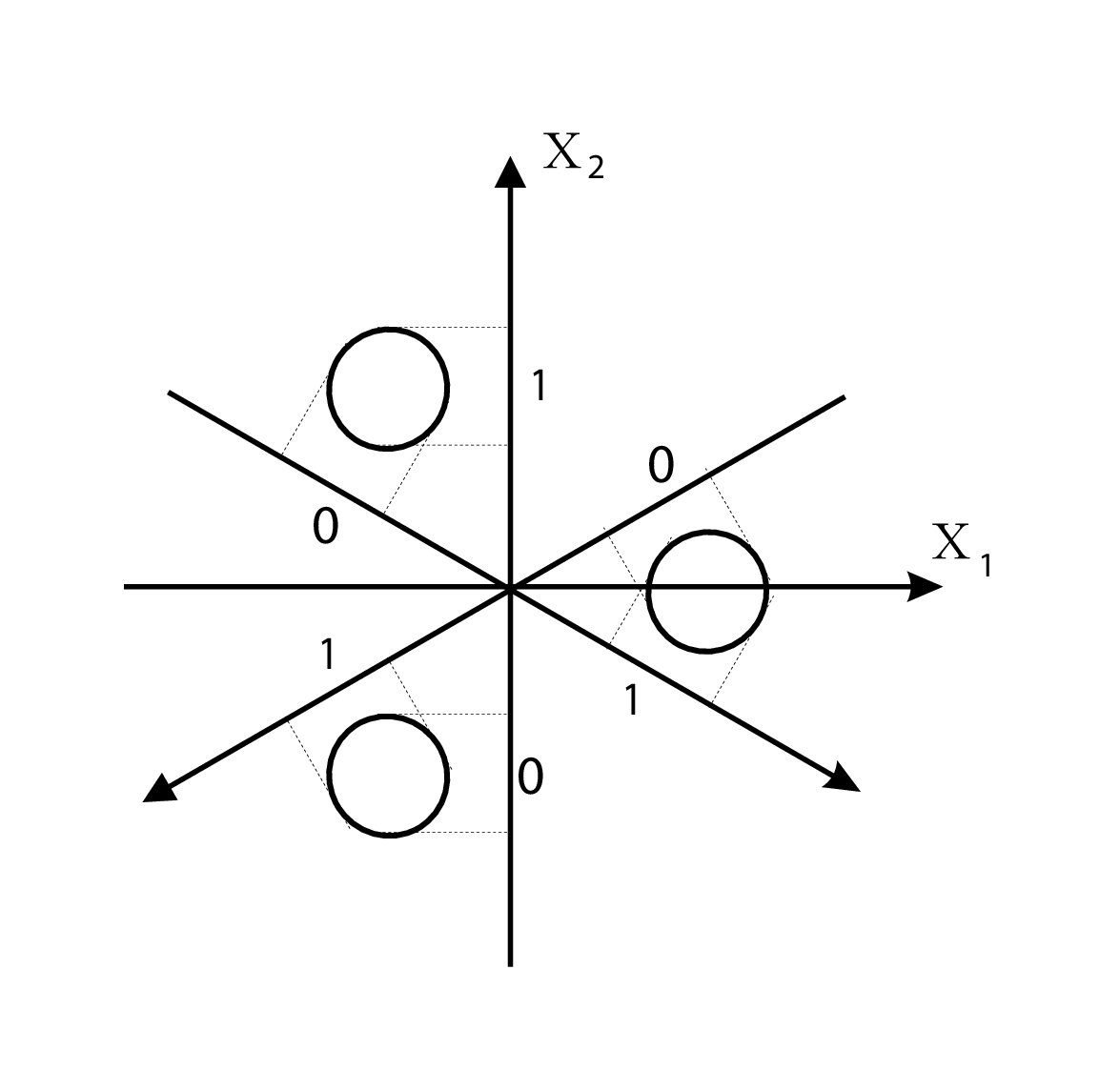}
\caption{Phase-space picture of the three-state protocol\label{ps-3}}\end{center}\end{figure}

The encoding of the three-state protocol is schematically described on the phasespace as in FIG. \ref{ps-3} (The ``1''s and ``0''s in FIG. \ref{ps-3} represent Alice's bit encoding associated with Bob's basis).  Alice sends the coherent state $| \alpha '  e^{i \phi_A }\rangle $ with $\phi_A = \{0, 2 \pi /3 ,4\pi /3 \}$ and $\alpha ' \equiv \frac{2}{\sqrt 3} \alpha$. 
Bob measures $\hat x (\phi_B)$ with $\phi_B= \{\pi /2 , - \pi /6, -5 \pi /6 \}$.  
We can easily see that if $( |\phi_A-\phi_B| \mod \pi )$ is different from $\pi /2$, then the quadrature distribution takes the form of the one in FIG. \ref{dist}. In such a condition, we can perform the postselection procedure. 
If $(|\phi_A-\phi_B| \mod \pi )=\pi /2$, the combination of $(\phi_A, \phi_B)$ is wrong basis and discarded.

Using Alice's bit encoding rule in Sec. II, we obtain Table II. From Table II, we can see that the wrong-basis case occurs with probability $1/3$. Thus, the efficiency is 
\begin{equation}P_e=\frac{2}{3},\end{equation} which is 4/3 times higher than that of the O4 protocol. 

\begingroup
\squeezetable
\begin{table}[hbtp]
\caption{Alice's bit encoding of the three-state protocol}
  \begin{tabular}{c||c|c|c|c|c|c|c|c|c|}
    \hline
    $\phi_A $ & 0 &0   &  0 &  $ 2\pi/ 3 $ &   $ 2 \pi/ 3 $ &   $ 2\pi/ 3 $ &  $ 4\pi/ 3 $ &    $ 4\pi/ 3 $ &   $ 4  \pi/ 3 $    \\
    \hline
     $\phi_B $  &  $ \pi/ 2 $  &  $- \pi /6 $  & $ -5 \pi / 6 $  &  $ \pi/ 2 $  &  $- \pi /6 $  & $ -5 \pi / 6 $ &  $ \pi/ 2 $  &  $- \pi /6 $  & $ -5 \pi / 6 $   \\
    \hline
     $\langle \hat x \rangle $  & 0 &  $\alpha $  & $- \alpha $   &  $\alpha $  &  $-\alpha $& 0 & $ - \alpha $ &  0 & $\alpha $  \\
    \hline
  $A$ &  & 1& 0& 1 & 0& & 0 & & 1 \\
    \hline
    \end{tabular}
\end{table}
\endgroup

This encoding is based on the idea that Alice applies different encoding according to Bob's choice of basis. In CV QKD protocols, an efficient encoding where two variables are encoded on two conjugate quadratures has already been common. Our proposal can be interpreted as a generalization of this idea, namely, we encode two values on non-conjugate quadratures. 

As a security aspect of the three-state protocol, it should be checked whether the minimum uncertainty is confirmed in the case Bob measures three different quadratures. Suppose that Bob observes $(\Delta x_2)^2 = \frac{1}{4}$ and $(\Delta x_\phi)^2 = \frac{1}{4}$. Then, from the definition of $\hat x (\phi)$,
\begin{eqnarray}
(\Delta x_\phi)^2 &\equiv & \langle (\hat x_1 \cos \phi + \hat x_2 \sin \phi )^2 -\langle \hat x_1 \cos \phi + \hat x_2 \sin \phi  \rangle^2  \rangle\nonumber
\\ &=& (\Delta x_1)^2 \cos^2 \phi + (\Delta x_2)^2 \sin^2 \phi  \nonumber \\
& &+  \left( \langle \hat x_1 \hat x_2 + \hat x_2  \hat x_1\rangle - 2  \langle \hat x_1    \rangle   \langle \hat x_2    \rangle \right) \sin \phi  \cos \phi ,
\end{eqnarray} and we obtain
\begin{eqnarray}
  \left( (\Delta x_1)^2 - \frac{1}{4} \right) +  \left(\langle \hat x_1 \hat x_2 + \hat x_2  \hat x_1\rangle - 2  \langle \hat x_1    \rangle   \langle \hat x_2    \rangle \right) \tan \phi  =0
.\end{eqnarray}
Since this relation holds for both $\phi = -\pi /6 $ and $\phi = -5\pi /6 $ in the three-state protocol, we obtain 
\begin{eqnarray}
    (\Delta x_1)^2 = \frac{1}{4} 
.\end{eqnarray}

\subsection{Efficient four-state protocols}
\begin{figure}[htbp]
\begin{center}
\includegraphics[width=5cm]{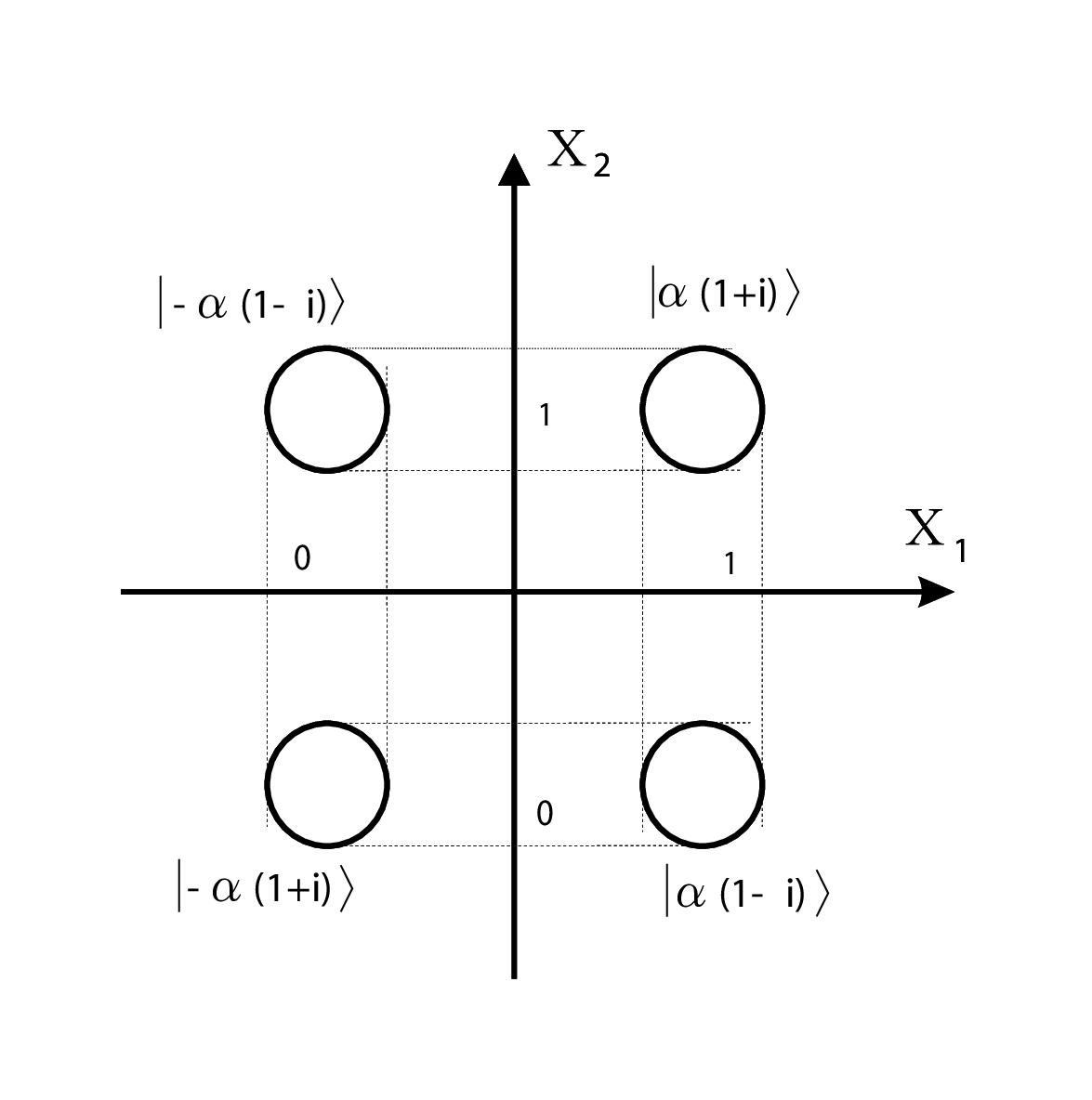}
\caption{Phase-space picture of the efficient four-state protocol\label{ps-e4}}\end{center}\end{figure}
The encoding of the efficient four-state (E4) protocol is schematically described in FIG. \ref{ps-e4}. Alice sends $| \alpha '' e^{i\phi_A} \rangle$ with $\phi_A = \{\pi /4, 3\pi /4, 5\pi /4, 7\pi /4 \}$ and $\alpha '' \equiv \sqrt 2 \alpha$. Bob measures $\hat x (\phi_B )$ with $\phi_B = \{0, \pi /2 \}$. After the transmission (i) Bob informs Alice of his phase. Then, (ii) Alice tells Bob that her preparation of the states belongs to which one of the four sets $\{| \alpha ''e^{i\pi/4} \rangle  , | \alpha ''e^{3i\pi/4}  \rangle \}$, $\{| \alpha'' e^{3i\pi/4} \rangle , |\alpha'' e^{i5\pi/4} \rangle   \}$, $\{| \alpha ''e^{5i\pi/4} \rangle , | \alpha ''e^{7i\pi/4}  \rangle   \}$, and $\{| \alpha ''e^{7i\pi/4} \rangle   ,| \alpha ''e^{i\pi/4} \rangle   \}$ according to Bob's choice of the phase. Namely, if $\phi_B = 0$, Alice tells Bob that the state belong to either $\{| \alpha ''e^{i\pi/4} \rangle  , | \alpha ''e^{3i\pi/4}  \rangle \}$ or $\{| \alpha ''e^{5i\pi/4} \rangle , | \alpha ''e^{7i\pi/4}  \rangle   \}$. If $\phi_B =  \pi /2 $, Alice tells Bob that the state belong to either $\{| \alpha'' e^{3i\pi/4} \rangle , |\alpha'' e^{i5\pi/4} \rangle   \}$ or $\{| \alpha ''e^{7i\pi/4} \rangle   ,| \alpha ''e^{i\pi/4} \rangle   \}$.

 In every case, the state takes the form similar to the one in FIG. \ref{dist} and we can perform the postselection procedure. Thus, the efficiency is \begin{equation}P_e =1.\end{equation} This is also clear from the configuration of the states in FIG. \ref{ps-e4}. Applying Alice's bit encoding rule we obtain Table III.  
\begingroup
\squeezetable
\begin{table}
\caption{Alice's bit encoding for the efficient four-state protocols}
  \begin{tabular}{c||c|c|c|c|c|c|c|c|}
    \hline
    $\phi_A $ & $\pi/ 4$ & $\pi/ 4$   &  $3\pi/ 4$  &  $3 \pi/ 4$ &   $5\pi/ 4$ &  $5 \pi/ 4$ & $7 \pi/ 4$ & $7 \pi/ 4$   \\
    \hline
     $\phi_B $  &  0  &  $\pi /2$  &  0  &  $\pi /2$ & 0 & $\pi /2$ & 0 & $\pi /2$ \\
    \hline
     $\langle \hat x \rangle $  &  $\alpha $  &  $\alpha $  & $- \alpha $   &  $\alpha $  &  $-\alpha $& $- \alpha $ & $\alpha $ &  $-\alpha $  \\
    \hline
    $A$ & 1 &1 &0& 1& 0 &0& 1& 0  \\
    \hline
  \end{tabular}
\end{table}
\endgroup

Now we present another four-state protocol by modifying the classical communication step of the E4 protocol. 
Let us assume that Alice omits the announcement (ii), then the protocol still works without any further modification and the information that Alice and Bob share does not change. This is because Bob's distribution and bit decoding do not depend on the announcement (ii). The information is encoded on the pairs of the states, and it is not necessary to identify the states in each pair. Therefore, the redundant state information need not be announced.

This modification possibly enhances the security because Eve cannot exploit the classical information (ii). The loss of the information can be described by using the terms of mixed states. From Eve's point of view, Alice's preparation of states is not in a set of pure coherent states but in a set of mixtures of coherent states. 
 We call this modified protocol the \textit{mixed state protocol based on the four states} (MB4).

We have seen that the postselection protocol can be demonstrated without wrong-basis signal as in other CV QKD protocols. Conversely, if there are wrong-basis signals in a phase encoding CV QKD as in the O4 protocol, the signals are supposed to have some useful information and may play a role of the decoy states. Namely, the wrong-basis signals restrict Eve's possible operations. Such aspect has already been pointed out in Refs. \cite{hirano,namiki1}. 

\subsection{Six-state protocols}
In a similar manner to the E4 and MB4 protocols, we can find a six-state protocol and its mixed-state version. 

\begin{figure}[hbtp]
\begin{center}
\includegraphics[width=5cm]{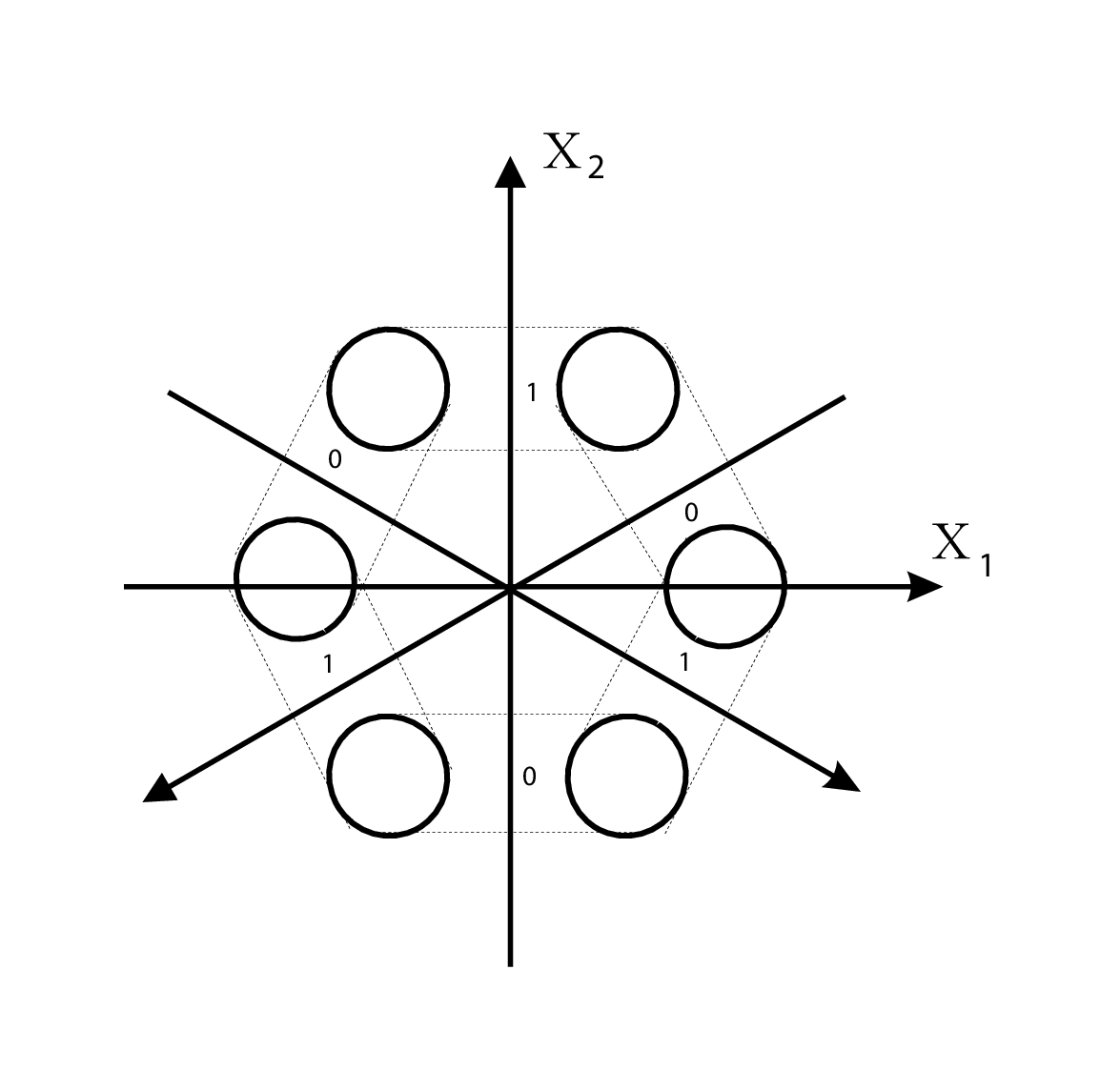}
\caption{ Phase-space picture of the six-state protocol \label{ps-6}}\end{center}
\end{figure}

The encoding of the six-state protocols can be schematically described on the phasespace as in FIG. \ref{ps-6}. Alice sends $| { \alpha '}  e^{i\phi_A}  \rangle$ with $\phi_A =  m \pi /3,  \ m=\{0,1,2, \cdots ,  5\}$. Bob measures $\hat x (\phi_B)$ with $\phi_B =  \{\pi /2, -\pi /6, -5\pi /6 \}$. 
The configuration of the six-state protocol in FIG. \ref{ps-6} has a similar symmetry to that of the three-state protocol in FIG. \ref{ps-3}. If we make a mirror reflection of the three states associated with any one of the measured axes on the phasespace, we can obtain the six-state configuration. 

From Fig. \ref{ps-6}, we can see that if $(| \phi_A- \phi_B| \mod \pi )$ is different from $\pi /2 $, the quadrature distribution takes the form similar to the one in Fig. \ref{dist}. In such a condition, we can perform the postselection procedure. The other cases, i.e., $(| \phi_A- \phi_B| \mod \pi )=\pi /2 $, are for wrong-basis result.

Using the bit encoding rule, we can obtain Table IV.  
From Table IV, we can see that the efficiency is \begin{equation}P_e=\frac{2}{3},\end{equation} which is equivalent to that of the three-state protocol.

\begin{widetext}
\begingroup
\squeezetable
\begin{table}[hbpt]
\caption{Alice's bit encoding of the six-state protocols}
  \begin{tabular}{c||c|c|c|c|c|c|c|c|c|c|c|c|c|c|c|c|c|c|}
    \hline
    $\phi_A $ & 0 &0   &  0 &  $ \pi/ 3 $ &   $  \pi/ 3 $ &   $ \pi/ 3 $ & $ 2\pi/ 3 $ &   $ 2 \pi/ 3 $ &   $ 2\pi/ 3 $ & $\pi$ &   $\pi$ &   $\pi$   & $ 4\pi/ 3 $ &    $ 4\pi/ 3 $ &   $ 4  \pi/ 3 $ & $ 5\pi/ 3 $ &   $ 5 \pi/ 3 $ &   $ 5\pi/ 3 $  \\
    \hline
     $\phi_B $  &  $ \pi/ 2 $  &  $- \pi /6 $  & $ -5 \pi / 6 $  &  $ \pi/ 2 $  &  $- \pi /6 $  & $ -5 \pi / 6 $ &  $ \pi/ 2 $  &  $- \pi /6 $  & $ -5 \pi / 6 $ &  $ \pi/ 2 $  &  $- \pi /6 $  & $ -5 \pi / 6 $  &  $ \pi/ 2 $  &  $- \pi /6 $  & $ -5 \pi / 6 $ &  $ \pi/ 2 $  &  $- \pi /6 $  & $ -5 \pi / 6 $  \\
    \hline
     $\langle \hat x \rangle $  & 0 &  $\alpha $  & $- \alpha $   &  $\alpha $  &  0 & $-\alpha $ & $  \alpha $ &  $-\alpha $ & 0 & 0 &  $-\alpha $  & $  \alpha $   &  $-\alpha $  &  0&$ \alpha $ & $ - \alpha $ &  $ \alpha $ & 0  \\
    \hline
    $A$ & & 1& 0& 1&  & 0 & 1& 0&  & & 0& 1&0&  &0 & 0& 1&    \\
    \hline
  \end{tabular}
\end{table}
\endgroup
\end{widetext}

The classical communication step in the six-state protocol is as follows: (i) Bob informs Alice of his phase $\phi_B$ and then (ii) Alice tells Bob that her preparation of the states is wrong basis or belongs to either two-state subset in the one of the three sets 
\begin{eqnarray}
\left\{ \{| \alpha ' e^{i\pi/3}  \rangle , | \alpha 'e^{5i\pi/3}  \rangle \}, \{| \alpha 'e^{2i\pi/3}  \rangle , | \alpha 'e^{4i\pi/3}  \rangle \} \right\},   \\
\left\{ \{| \alpha  '\rangle , | \alpha 'e^{2i\pi/3}  \rangle \}, \{| \alpha 'e^{i\pi}  \rangle , | \alpha 'e^{5i\pi/3}  \rangle \} \right\},\end{eqnarray} and
\begin{eqnarray}
\left\{ \{| \alpha 'e^{i\pi/3}  \rangle , | \alpha 'e^{i\pi}  \rangle \}, \{| \alpha '  \rangle , | \alpha 'e^{4i\pi/3}  \rangle \} \right\}, \label{6set}
\end{eqnarray}
according to the value of $\phi_B = \pi /2 , - \pi /6$, and $ -5 \pi /6$ , respectively.

Let us modify the step (ii) as (ii)$'$ Alice tells Bob that her preparation of the states is wrong basis or not.  Then, the protocol still works and the performance is essentially the same. This provides the \textit{mixed state protocol based on the six states} (MB6). In this protocol, Alice's preparation of states is considered to be one of the equal-probability mixtures of the states in one of the subsets. 

Note that all above protocols can be performed essentially in the same experimental setup based on the interferometer because the modifications are given in the way of phase modulation and classical communication steps.   

\subsection{Eight-state protocols}
We describe an eight-state protocol and its mixed-state version.
Different from the previous ones, this protocol requires not only the phase modulation, but also the amplitude modulation in the interferometric implementation.

\begin{figure}[here]
\begin{center}
\includegraphics[width=5cm]{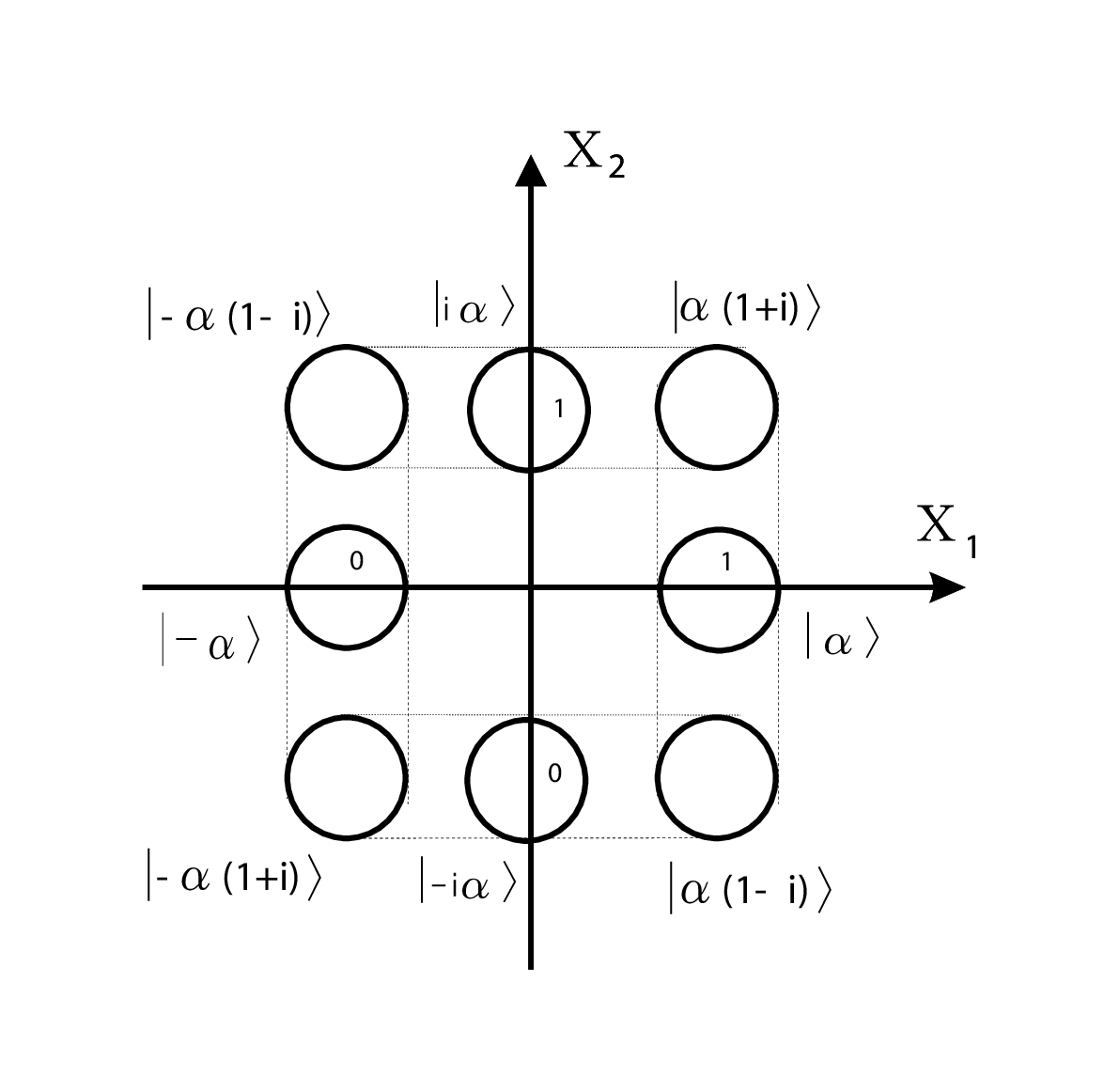}
\caption{Phase-space picture of the eight-state protocol \label{ps-8}}\end{center}\end{figure}

The encoding of the eight-state protocols can be schematically described on the phasespace as in FIG. \ref{ps-8}. Alice sends coherent state $\{| \alpha e^{i \frac{m}{2}\pi}\rangle , | \alpha '' e^{i \frac{2m'+1}{4}\pi}\rangle \}$ by choosing $\phi_A =  m \pi /2,    m=\{0,1,2,3\}$ with $  \alpha $ and $\phi_A =  (2m'+1) \pi /4,    m'=\{ 0,1,2,3\}$ with $ \alpha ''  = \sqrt 2 \alpha $. Bob chooses his phase $\phi_B =  \{0, \pi /2  \}$. 

From FIG. \ref{ps-8}, we can see that if $(| \phi_A- \phi_B| \mod \pi )$ is different from $\pi /2 $, the quadrature distribution takes the form similar to the one in FIG. 2. In such a condition, we can perform the postselection procedure. The other cases, i.e., $(| \phi_A- \phi_B| \mod \pi ) =\pi /2 $, are for wrong-basis result. Using the bit encoding rule, we can obtain Table V.  
From Table V, we can see that the efficiency is \begin{equation}P_e =\frac{3}{4}\end{equation} which is higher than that of the O4 protocol by the factor of $3/2$. 

\begin{widetext}
\begingroup
\squeezetable
\begin{table}[thbp]
    \caption{Alice's bit encoding of the eight-state protocol}
  \begin{tabular}{c||c|c|c|c|c|c|c|c|c|c|c|c|c|c|c|c|c|c|}
    \hline
    $\phi_A $ & 0 &0   &  $\pi / 4$ &  $ \pi/ 4 $ &   $  \pi/ 2 $ &   $ \pi/ 2 $ & $ 3\pi/ 4 $ &   $ 3 \pi/ 4 $ &   $  \pi  $ & $\pi$ &   $5\pi /4 $ &   $5 \pi /4$   & $ 3\pi /2 $ &    $ 3\pi/ 2 $ &   $ 7  \pi/ 4 $ & $ 7\pi/ 4 $   \\
    \hline
     $\phi_B $  &  0  &  $ \pi/ 2 $  &0  &  $ \pi/ 2 $  &0  &  $ \pi/ 2 $  &0  &  $ \pi/ 2 $  &0  &  $ \pi/ 2 $  &0  &  $ \pi/ 2 $  &0  &  $ \pi/ 2 $  &0  &  $ \pi/ 2 $    \\
    \hline
     $\langle \hat x \rangle $  & $\alpha $ &  0  & $\alpha $   &  $\alpha $  & 0 & $\alpha $ & $ - \alpha $ &  $\alpha $ & $\alpha $ & $- \alpha $ &  $0 $  & $- \alpha $   &  $-\alpha $  &  0& $ \alpha $ & $   -\alpha $    \\
    \hline 
     $A$ &1 & &1&1& &1&0&1&0 &0& &0&0 & &1&0 \\
    \hline
  \end{tabular}
\end{table}\endgroup
\end{widetext}

In the classical communication step, (i) Bob informs Alice of his phase $\phi_B$ and then (ii) Alice tells Bob that her preparation of the states is wrong basis or belongs to which one of the two-state subsets in the two sets 
\begin{widetext}
\begin{eqnarray}
\left\{ \{| \alpha   \rangle , | \alpha e^{i\pi }  \rangle \} , \{| \alpha '' e^{i\pi/4}  \rangle, | \alpha  ''e^{3i\pi/4}  \rangle \}, \{| \alpha  ''e^{5i\pi/4}  \rangle , | \alpha  ''e^{7i\pi/4}  \rangle \} \right\},\end{eqnarray} and
   \begin{eqnarray}
 \left\{ \{| \alpha e^{i\pi/2}  \rangle , | \alpha e^{3i\pi  /2}  \rangle \} , \{| \alpha  ''e^{i\pi/4}  \rangle, | \alpha  ''e^{7i\pi/4}  \rangle \}, \{| \alpha  ''e^{3i\pi/4}  \rangle , | \alpha  ''e^{5i\pi/4}  \rangle \} \right\}, \label{8set}
\end{eqnarray} \end{widetext}
according to the value of $\phi_B = 0 ,$ and $\pi /2$ , respectively. 

 Let us modify the process (ii) as (ii)$'$ Alice tells Bob that her preparation of the states is wrong basis or not. Then, still the protocol works and the performance is essentially the same. In this case, Alice's preparation of states is considered to be a three-state equal-probability mixture. We call this modified eight-state protocol the \textit{mixed state protocol based on the eight states} (MB8).
The MB8 protocol is considered to be a mixture of the O4 and E4
protocols with certain changes in the classical communication step. Thus, it is possible to switch the protocols by changing the classical communication after the transmission.

\subsection{Generalization of protocols}
We present a generalized mixed-state protocol which includes the MB4 and MB8 protocols. This generalization may
not have practical utility but it is useful for the discussion of the security (see Sec. IV).

\begin{figure}[hbtp]
\begin{center}
\includegraphics[width=4cm]{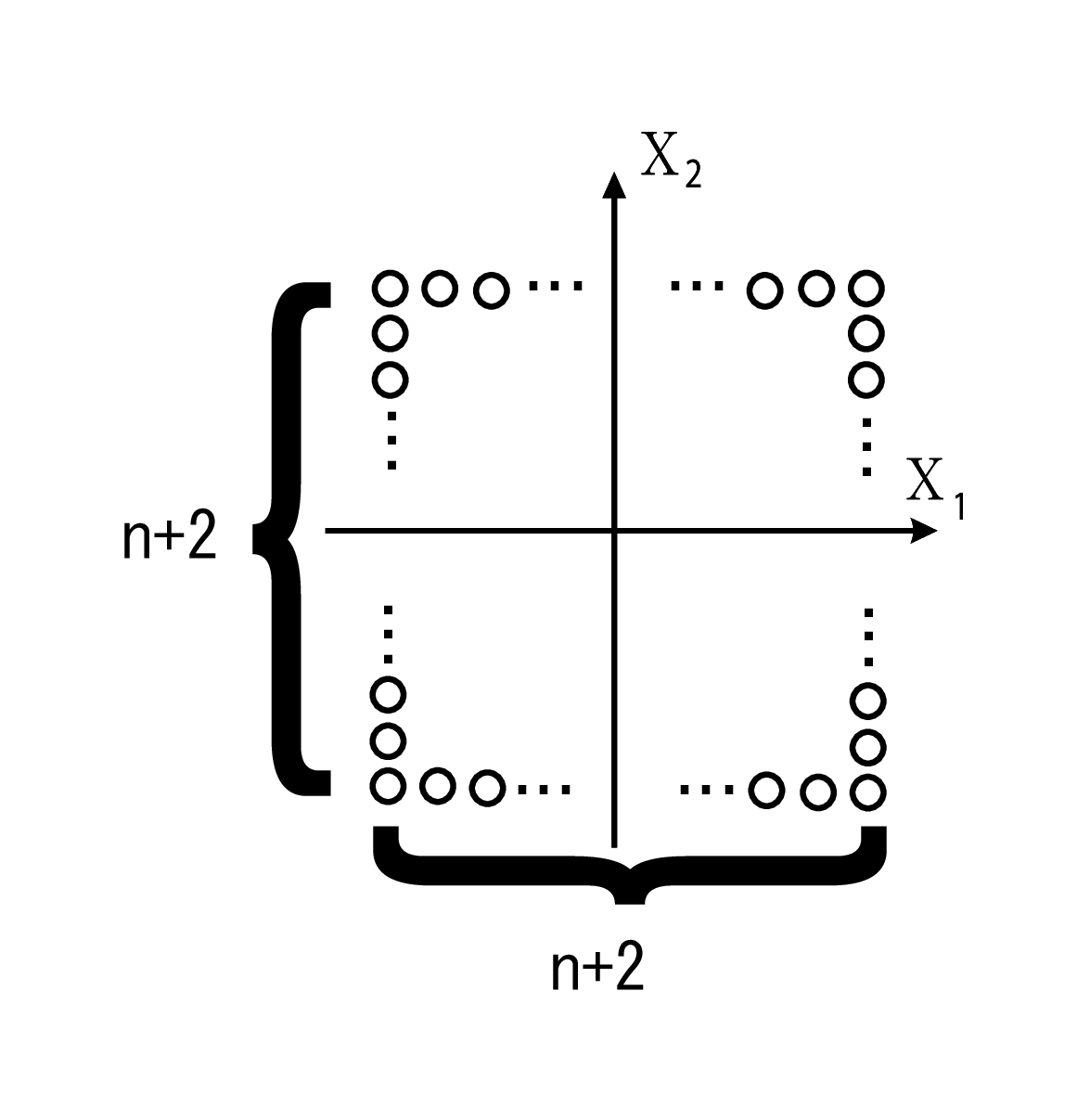}
\caption{Phase-space picture of the $4n +4$-state protocol \label{ps-4n}}\end{center}\end{figure}

The encoding of the generalized protocol can be schematically described on the phasespace as in FIG. \ref{ps-4n}. Alice randomly sends one of the $4n +4$ states, $| \alpha (\pm 1 +i (\frac{2k}{n+1} -1 ) ) \rangle $, $| \alpha (\pm i + (\frac{2m}{n+1} -1 ) ) \rangle $, $k= 0,1,2, \cdots , n+1$, $m= 1,2, \cdots , n$. Bob measures one of the quadratures, $\hat x_1$ or $\hat x_2$. 

Alice encodes bit ``1'' for $| \alpha (  1 +i (\frac{2k}{n+1} -1 ) ) \rangle $ and ``0'' for $| \alpha (- 1 +i (\frac{2k}{n+1} -1 ) ) \rangle $ in the case that Bob measures $\hat x_1$. Alice encodes bit ``1'' for $|  \alpha ( i +  (\frac{2k}{n+1} -1 ) ) \rangle $ and ``0'' for $| \alpha (-i +  (\frac{2k}{n+1} -1 ) ) \rangle $ in the case that Bob measures $\hat x_2$. The other cases are discarded. In this protocol, Alice's preparation of states is considered to be a $n+2$-state equal-probability mixture.
We can see that $n=0$ corresponds to the MB4 protocol and $n=1$ correspond to the MB8 protocol.
The efficiency is given by
\begin{eqnarray}
P_e = \frac{2+n}{2+2 n}.
\end{eqnarray}

\section{Security against individual beam splitting attack}
\subsection{Individual beam splitting attack}

The safety of the QKD protocols under lossy channel is estimated by assuming the beam splitting attack where Eve replaces the lossy channel with lossless one and splits the portion of signal corresponds to the loss. Here we consider the individual beam splitting attack where Eve stores the portion for arbitrarily long time with a perfect quantum memory and she performs her measurement on individual signal independently after she learns Bob's basis of each signal.

In the present protocols, Alice's preparation of states is represented by the form of the coherent-state mixture  
\begin{eqnarray}
\hat \rho (p_i, \alpha_i) =\sum_{i} p_i | \alpha_i \rangle \langle \alpha_i |
\end{eqnarray}
The states of Bob and Eve under the beam splitting attack are given by $\hat \rho (p_i, \sqrt \eta \alpha_i) $ and $\hat \rho (p_i, \sqrt{1- \eta} \alpha_i) $ respectively,
where $\eta$ is the line transmission ($0 < \eta \le 1$).

\subsection{Eve's knowledge}
Let us assume the situation that Eve receives one of the binary states $\hat \rho_\pm$ whose subscript corresponds to Alice's bit encoding and each of them appears with the equal probability, $\frac{1}{2}$. An upperbound of Eve's knowledge represented by the fraction of the bit sequence deleted in the privacy amplification \cite{lut96,lut99,lut00} is given by 
\begin{eqnarray}
\tau_u &=& \log_2 (2- | \langle\psi_+ | \psi_- \rangle |^2) ,
\end{eqnarray} where $| \psi_\pm \rangle$ is a purification of $\hat \rho_\pm$. This formula is originally given for the pure-state signal \cite{A57}. For the case of the mixed-state signal, by assuming Eve obtains a purification that always includes the original mixed-state signal, we can safely estimate an upperbound.   
A tight estimation of the inner product is directly calculated by the fidelity \cite{NC00}
\begin{eqnarray} 
F(\hat \rho_+, \hat \rho_- )  &=& \max_{    |\psi_\pm\rangle } |\langle \psi_+ | \psi_- \rangle |  =\textrm{Tr} \sqrt{\sqrt {\hat \rho_+ }\hat \rho_- \sqrt {\hat \rho_+ } }
\end{eqnarray}
where the maximization is taken over all the purifications.

In the following we determine Eve's states under the beam splitting attack for each of the protocols and estimate the inner product $| \langle\psi_+ | \psi_- \rangle |$. It is shown that all of them have the same bound of the inner product $| \langle\psi_+ | \psi_- \rangle |= e^{-2 (1- \eta )\alpha ^2}$, which gives the same upperbound of Eve's knowledge. 

For the three-state protocol, Eve's states are given by
\begin{eqnarray}
\hat \rho_\pm =  | \sqrt{1- \eta} \alpha 'e^{\pm \frac{2}{3} \pi}\rangle \langle \sqrt{1- \eta}\alpha 'e^{\pm \frac{2}{3} \pi} |   \label{ro1}
\end{eqnarray} assuming that Bob set $\phi_B = \pi / 2$. Then, taking $| \psi_\pm \rangle = | \sqrt{1- \eta}\alpha 'e^{\pm \frac{2}{3} \pi} \rangle$, we obtain $| \langle\psi_+ | \psi_- \rangle |= 
e^{-2 (1- \eta ) \alpha ^2 }$. 
The same bound can be applied for $\phi_B = -\pi/6$ and $\phi_B = -5\pi/6$ because in that case $\hat \rho _\pm $ can be obtained from those in Eq. (\ref{ro1}) with proper rotations in the phasespace.
Similarly we omit the discussions of phase-covariant cases in the followings. Since the configuration of the six-state protocols can be obtained from that of the three-state protocol by the mirror reflection. The same bound can be applied to the pure version of the six-state protocols. 
 
For the E4 protocol, Eve's states are given by
\begin{eqnarray}
\hat \rho_\pm &=&  |  \sqrt{1- \eta} (1\pm i )\alpha  \rangle \langle  \sqrt{1- \eta}(1\pm i )\alpha   |
\end{eqnarray} assuming that Bob set $\phi_B = \pi /2 $ and Alice announced her preparation was in the pair $|  (1 \pm i )\alpha  \rangle  $. Then, taking $| \psi_\pm    \rangle =|  (1 \pm i )\alpha  \rangle $ we obtain the same bound, $| \langle\psi_+ | \psi_- \rangle |=e^{-2 (1- \eta ) \alpha ^2 }$.

For the MB4 protocol, Eve's states are given by
\begin{eqnarray}
\hat \rho_\pm &=&\frac{1}{2} \Big (  |\pm \sqrt{1- \eta} (1+i )\alpha  \rangle \langle \pm \sqrt{1- \eta}(1+i )\alpha   | \nonumber \\ & & + |\pm \sqrt{1- \eta} (1-i ) \alpha \rangle \langle \pm \sqrt{1- \eta} (1-i ) \alpha  | \Big )
\end{eqnarray} assuming that Bob set $\phi_B = 0$. Using the formula of the fidelity in Appendix \textbf{A}, we obtain the estimation $F = e^{-2 (1- \eta ) \alpha ^2 }$.
 Since Eve's states of the pure version of the eight-state protocols is that of either the O4 protocol or the E4 protocol. Therefore, the pure version of the eight-state protocols also has the same bound.

For the MB6 protocol, Eve's states are given by
\begin{eqnarray}
\hat \rho_\pm &=& \frac{1}{2} \Bigg(  | \sqrt{1- \eta } \alpha 'e^{\pm \frac{\pi}{3} } \rangle \langle  \sqrt{1- \eta} \alpha 'e^{\pm \frac{\pi}{3}  } | \nonumber \\ & & + | \sqrt{1- \eta} \alpha 'e^{\pm \frac{2}{3} \pi}  \rangle \langle \sqrt{1- \eta} \alpha 'e^{\pm \frac{2}{3} \pi}   | \Bigg)
\end{eqnarray} assuming that Bob set $\phi_B = \pi / 2$. Then, using the formula of the fidelity in Appendix \textbf{A} again, we obtain the estimation  $F = e^{-2 (1- \eta ) \alpha ^2 }$.

For the generalized protocol including the MB4 and MB8 protocol, Eve's states are given by  
\begin{eqnarray}
\hat \rho_ \pm &\equiv &\frac{1}{ {n+2}}\sum_{k=0}^{n+1} | \beta_\pm ( n, k)\rangle \langle \beta_\pm ( n, k)|,  \\
\beta_\pm (n,k) &\equiv&  \sqrt{1-\eta } \alpha \left\{ \pm 1 +i \left(\frac{2k-n-1}{n+1}  \right) \right\} ,
\end{eqnarray}
assuming Bob set $\phi_B = 0 $. 
We can see that the following states are purifications of $\hat \rho_ \pm$,
\begin{eqnarray}
| \psi_ \pm \rangle &\equiv &\frac{1}{\sqrt{n+2}}\sum_{k=0}^{n+1} | \beta_\pm (n, k)\rangle | k\rangle e^{\pm i \omega (k)}, \label{puripu}\\
&&\omega (k ) \equiv (1- \eta)  \alpha ^2  \left(\frac{2k-n-1}{n+1}  \right),
\end{eqnarray}
where $\{ | k\rangle \}$ is an orthonormal basis set of an extended space. The bound is given by 
\begin{eqnarray}
|\langle \psi_+ | \psi_ - \rangle |&=& e^{-2 (1- \eta )\alpha ^2}
\end{eqnarray} independent of $n$.

 The purification $| \psi _\pm \rangle $ of Eq. (\ref{puripu})  implies that Eve knows the extra-state information $k$ which is never announced in the mixed-state protocols. Given this information Eve's problem is just to distinguish the two pure states $|\beta_\pm (n,k) \rangle$. Thus, the bound is equal to that of the pure-state case.   
In this sense, the information from the mixed-state signal is upper bounded by that of the pure-state case and the omission of the classical communication possibly reduces Eve's information.
This fact suggests that mixed-state protocols are advantageous.
However, optimization problems for mixed-state signals are technically difficult compared with those for pure-state cases in general.
For the pure-state protocols, the upperbound is achievable by the optimal measurement which minimizes the error rate. On the other hand, it is not sure that the upperbound can be tight for the mixed-state protocols. If the bound is not tight, it implies that the mixed-state protocol is more secure.

\subsection{Formula of secure key gain}

The secure key gain \cite{lut96,lut99,lut00} as a estimation of QKD performance 
against individual BS attack for the present protocols is given by 
\begin{eqnarray} 
G(\alpha,\eta,x_0)&=& P_{e} \left( \sum_{x} P_B(x,\sqrt \eta \alpha ) i(q_B(|x|,\sqrt \eta \alpha)) -\tau_u \right).
\end{eqnarray}
where 
\begin{equation}
 i(q) \equiv 1+ q \log _2 q + (1-q) \log _2 (1- q)
\end{equation} is the mutual information of the binary symmetric channel, and 
 $\tau_u = \log_2 (2- e^{-4 (1- \eta )\alpha ^2})$ is given from the previous subsection. 
Since the correct-basis distributions, the BER, and $\tau$ are the same as those of the O4 protocol, the value of the gain is the same as that of the O4 protocol \cite{namiki2} except for the factor $P_e$.

In the presence of noise, the estimation of the gain is in progress. As long as we use the coherent states and homodyne detection the limitation of all the protocols can be found \cite{namiki2,namiki3}. 

\section{conclusion and remarks}
We have proposed several phase-encoding protocols for quantum key distribution using coherent states and postselection. 
The modified phase encodings reduce the probability of wrong basis and increase the efficiency. 
The proposed protocols include the mixed-state protocols, those are obtained from the protocols exploiting more than three states by omitting the annoucement of the redundant state information in the classical communication steps.


 We have investigated the security of the protocols against the individual beam splitting attack. We showed that the improvement of the key gains is simply proportional to the efficiency and no substantial difference is observed whether the protocol is mixed-state version or not. 
This result is depending on the way of our analysis and it leaves an open question whether the modified protocols provide physically different condition in the security of QKD particularly on the relation with the introduction of the mixed states. 

There exist several possibilities to make other protocols. Trivial one is to increase the way of phase modulations.   
To use asymmetric configuration or biased choice of basis \cite{biase} may be interesting. Extensive search of protocols and optimization of efficiencies are left for future works.


\appendix
\section{fidelity between the mixtures of two coherent states}
\begin{figure}[htb]
\begin{center}
\includegraphics[width=6cm]{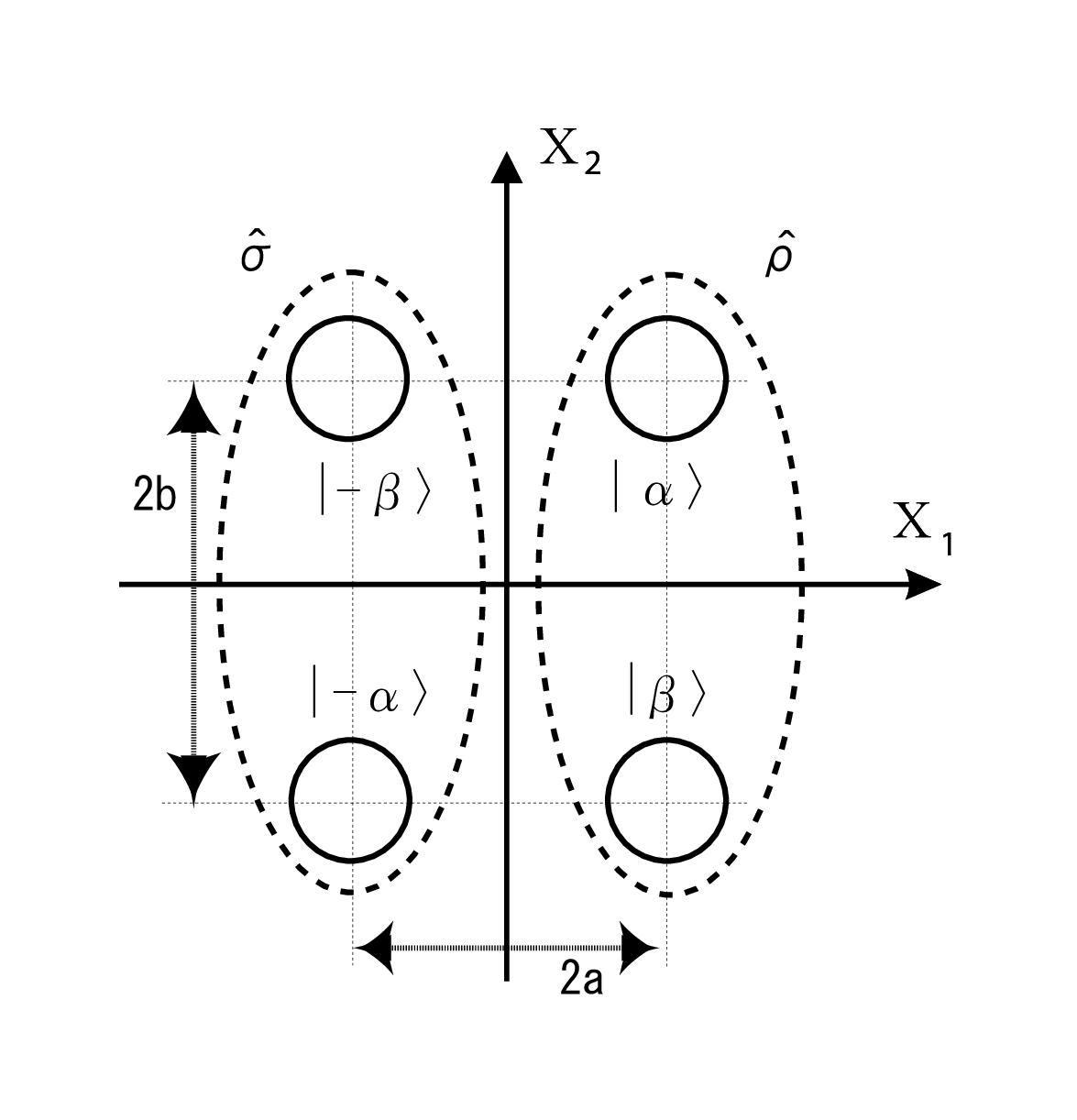}
\caption{Phase-space picture of the mixtures of two coherent states\label{ps-f}}\end{center}
\end{figure}

We derive a formula of the fidelity between the mixtures of two coherent states 
\begin{eqnarray}
\hat \rho  &=& \frac{1}{2}\left( | \alpha \rangle  \langle  \alpha  |+ |   \beta \rangle  \langle  \beta |\right) \\
\hat \sigma &=& \frac{1}{2}\left( | -\alpha  \rangle  \langle  -\alpha   |+ |  -\beta    \rangle  \langle -\beta  |\right) 
\end{eqnarray} with $\alpha = a+ ib , \beta = a -ib ,  (a, b\ge 0) $. A phase-space configuration of the states is shown in FIG. \ref{ps-f}.
Since we can write\begin{eqnarray} 
           \sqrt {\hat \rho }&=& \frac{1}{\sqrt 2}\left( \sqrt{1+\gamma} | + \rangle  \langle + |+\sqrt{1-\gamma} |  - \rangle  \langle - |\right)
\end{eqnarray} where we defined
an orthonormal basis which diagonalizes $\hat \rho$ \begin{eqnarray}
|\pm \rangle &\equiv & \frac{| \alpha \rangle \pm e^{i \phi}| \beta \rangle }{\sqrt{2 (1\pm \gamma)}} \\
\gamma e^{i\phi}  & \equiv&\langle\beta  |  \alpha  \rangle, \gamma \ge 0  ,  
 \end{eqnarray} \begin{widetext}
we can find a matrix representation of $\sqrt {\hat \rho }  \hat \sigma \sqrt {\hat \rho } $ as 
\begin{eqnarray}
 \left ( \begin{array}{c}
\langle +| \\
\langle -| \end{array} \right)
\sqrt {\hat \rho } \hat \sigma \sqrt {\hat \rho }
 \left ( \begin{array}{cc}|+ \rangle  & |-\rangle 
 \end{array} \right)
 &\equiv & \frac{1}{4}  \left ( \begin{array}{cc}
 (1+ \gamma) \langle + | \hat \sigma|+ \rangle
  & \sqrt{1- \gamma^2} \langle + | \hat \sigma|- \rangle
\\
\sqrt{1- \gamma^2} \langle - | \hat \sigma |+ \rangle
 &(1- \gamma) \langle - | \hat \sigma|- \rangle
 \end{array} \right) \nonumber \\
 &=& \frac{e^{-4 a^2}}{4} \left ( \begin{array}{cc}
1+e^{-4 b^2} + 2 e^{-2 b^2} \cos  (4ab) 
& -2ie^{-2b^2} \sin (4 ab) 
\\
 2ie^{-2b^2} \sin (4 ab)
&1+e^{-4 b^2} - 2 e^{-2 b^2} \cos  (4ab) 
 \end{array} \right).
 \end{eqnarray}
Then, using the relation $\textrm{Tr} \sqrt{X} = \sqrt{\textrm{Tr} X + 2\sqrt {\det X}} $ for a 2$\times$2 positive matrix $X$, we obtain the fidelity
 \begin{eqnarray}
 F ({\hat \rho },\hat \sigma ) \equiv \textrm{Tr} \sqrt{\sqrt {\hat \rho }\hat \sigma \sqrt {\hat \rho } }= \sqrt{ \textrm{Tr}(\sqrt {\hat \rho }\hat \sigma \sqrt {\hat \rho })+2 \sqrt {\det  (\sqrt {\hat \rho }\hat \sigma \sqrt {\hat \rho }) }}
= e^{ -2 a ^2}. 
 \end{eqnarray}
 \end{widetext} 
 

\end{document}